\documentclass[
  a4paper, 
  12pt, onecolumn,
]{article}

\title{\bf     Hidden symmetries in the asymmetric exclusion process  
}
\author{        O. Golinelli, K. Mallick
\bigskip
\\ \ad          Service de Physique Th\'eorique, 
\\ \ad          Cea Saclay, 91191 Gif-sur-Yvette, France
}
\date{\normalsize  
               arXiv:cond-mat/0412353 - JSTAT (2004) P12001 
\\             Received 30 September 2004
\\             Published 2 December 2004
\\            
}

\newcommand  {\ad}{\normalsize\em}      
\begin{document}
\maketitle

\begin{abstract}
\normalsize
 
 We present a spectral study of the evolution matrix of the totally
 asymmetric exclusion process on a ring at half filling.  The natural
 symmetries (translation, charge conjugation combined with reflection)
 predict only two fold degeneracies. However, we have found that
 degeneracies of higher order also exist and, as the system size
 increases, higher and higher orders appear.  These degeneracies
 become generic in the limit of very large systems. This behaviour can
 be explained by the Bethe Ansatz and suggests the presence of hidden
 symmetries in the model.

\medskip \noindent 
 Keywords: ASEP, Markov matrix, symmetries, spectral degeneracies, 
 Bethe Ansatz.

\end{abstract}

\section{Introduction}

   The asymmetric simple  exclusion process (ASEP)   that 
 plays the role of a paradigm in non-equilibrium
 statistical physics   is  a  model 
 of  driven diffusive  particles on a lattice  interacting  through
 hard-core exclusion (Katz, Lebowitz and Spohn 1984).
 Introduced   originally    as a model for
 protein synthesis on RNA   (MacDonald and Gibbs 1969) 
  and for  hopping conductivity (Richards 1977) in
 solid electrolytes, the ASEP  has found a wide spectrum
 of  applications ranging from  traffic
 flow  
 (Chowdhury {\it et al.} 2000)
 to  surface growth 
 (Halpin-Healy and Zhang 1995, Krug 1997) 
  and sequence alignment (Bundschuh 2002). The ASEP
 has also motivated many theoretical studies,  providing 
 a `minimal'  model  sophisticated  enough to yield  a rich phenomenology but 
 elementary  enough to allow   exact  solutions 
[for a review see (Derrida 1998) and (Schutz 2001)].

 The ASEP is a memoryless stochastic   process governed by simple 
 local update rules that can be embodied in a  Markov evolution
 matrix $M$. This matrix contains all the dynamical  information 
 about the system:  for example,  the eigenvector of $M$ 
  with eigenvalue 0 
 represents   the steady state of the ASEP and  
 embodies  the stationary correlation functions. In particular, 
  the matrix product  representation  
 of  the stationary state (Derrida {\it et al.} 1993)
 allows to calculate steady state properties and
 large deviation functionals (Derrida {\it et al.} 2003). 
 More generally, the  spectrum of $M$ encodes  the relaxation behaviour
 of the ASEP and  all  non-stationary correlation functions.
 The Markov matrix $M$ can be diagonalized  with the 
  Bethe Ansatz  that  allows   to calculate 
  the energy gap  and   the dynamical exponent
 of the ASEP 
 (Dhar 1987,  Gwa and Spohn 1992, Kim 1995, Golinelli and Mallick 2004).

  Although  the steady state of the ASEP 
  and the lowest excitations  have been extensively studied, little effort
   has been devoted to investigate 
  global spectral  properties of  the  Markov matrix  
   (Bilstein and  Wehefritz  1997,
   Dud\-zi\'nski and Sch\"utz  2000,
   Nagy {\it et al.} 2002). 
  In this work,  we  focus  on the   degeneracies in the spectrum 
  of   the  Markov matrix  for the 
  totally asymmetric simple exclusion process (TASEP)
 on a  periodic ring  at half filling ({\it i.e.},   the density
 of particles is one-half).
 The exclusion  process   has some
   natural  invariance  properties by translation, charge conjugation
 and  reflection. At    half filling   these  symmetries 
 imply   the existence of doublets in the spectrum of $M$.
 However, we find that  multiplets with  degeneracies of higher order 
 also   exist for the TASEP.  Moreover, 
 the number of these multiplets and  their  degeneracies 
 are  given by simple combinatorial 
 factors. These remarkable  spectral properties suggest
 that the TASEP at half filling  possesses  some hidden symmetries
 other than the obvious ones.

  This article is organized as follows.  
 In section 2, we recall the definition of the ASEP on a periodic ring 
 and discuss  the translation, the  charge conjugation and  the reflection
 symmetries,  and their effect on the spectrum of the Markov matrix $M$.
 In section 3, we  present some numerical evidence for spectral
 degeneracies in the TASEP at half filling. We show that 
 the combinatorial factors involved can be understood, heuristically,
 from the Bethe Ansatz. In  
 Section 4, we present some  concluding remarks and discuss
  spectral  degeneracies in closely related models.
   In the Appendix, we analyze 
 the natural symmetries of the ASEP with a tagged particle.

\section{Natural symmetries of the ASEP on a ring}

\subsection{The model}

We consider the exclusion process   on a periodic
one dimensional lattice with $L$ sites (sites $i$ and $L + i$ are
identical) and $n$ particles. 
 A  lattice site cannot be  occupied by more than 
  one particle  ({\em exclusion rule}).
 The state of  a site  $i$ ($1 \le i \le L$)  can thus be characterized
 by  the Boolean number $\tau_i = 0, 1$ according as the site  $i$ is
 empty or occupied.

 The system evolves  with  time according to 
the following stochastic rule: a particle on a site $i$ at time $t$ jumps, in
the interval between  $t$ and $t+dt$, with probability $p\ dt$ to the
neighbouring site $i+1$ if this site is empty 
 and with  probability $q\ dt$ to the
 site $i-1$ if this site is empty. The jump  rates $p$ and $q$
 are normalized such that  $ p + q =1$. In 
  the totally asymmetric exclusion process (TASEP) 
 the jumps are totally biased in one direction ($p =1$ and $q =0$).

The  number $n$ of particles is conserved by the dynamics. The total number of
configurations for $n$ particles on a ring with $L$ sites is given by
$\Omega = L! / [ n! (L-n)!]$. 
A configuration $\mathcal{C}$  can be represented  by 
 the sequence $(\tau_1, \tau_2, \ldots, \tau_L).$
  We call $\psi_t(\mathcal{C})$
 the probability of configuration $\mathcal{C}$ at time $t$.
  As the exclusion process is a continuous-time Markov process, the time
 evolution of $\psi_t(\mathcal{C})$ is determined by the master equation 
\begin{equation}
    \frac{d}{dt} \psi_t(\mathcal{C})  = \sum_{\mathcal{C}'}
      M(\mathcal{C},\mathcal{C}') \psi_t(\mathcal{C}')  \, . 
\end{equation}
 The   Markov  matrix  $M$  encodes  the dynamics of the exclusion process:  
 the element  $ M(\mathcal{C},\mathcal{C}')$
 is the  transition rate  from configuration $\mathcal{C}'$ to $\mathcal{C}$
 and  the diagonal term $M(\mathcal{C},\mathcal{C}) =
  -  \sum_{\mathcal{C}'}  M(\mathcal{C}',\mathcal{C})$ 
represents  the exit rate
 from configuration $\mathcal{C}$.

 A right eigenvector $\psi$ is associated with the eigenvalue $E$
of $M$ if
\begin{equation}
  M \psi = E\psi      \, . 
  \label{eq:mpsi=epsi}
\end{equation}

The   matrix $M$ is a real non-symmetric matrix and, therefore, its
eigenvalues (and eigenvectors) are either real numbers or complex
conjugate pairs. The spectrum of $M$ contains the eigenvalue $E = 0$ and
the associated right eigenvector is the stationary state. For the ASEP
 on a ring the steady state    is uniform:
 the stationary probability of  any configuration 
  $\mathcal{C}$   is given by 
 $\psi(\mathcal{C}) = 1/\Omega$.
  Because the dynamics is ergodic ({\it i.e.}, $M$ is 
irreducible and aperiodic), the Perron-Frobenius theorem
 (see, for example, Gantmacher 1964)
implies that 
 0 is a non-degenerate eigenvalue and that all other 
eigenvalues $E$  have a strictly negative real
part; the relaxation time of the corresponding eigenmode is $\tau =
-1/\mathrm{Re}(E)$. 
\hfill\break 
 {\it Remark:} A configuration can  also be characterized
  by the positions of the $n$ particles
on the ring, $(x_1, x_2, \dots, x_n)$ with $1 \le x_1 < x_2 < \dots < x_n
\le L$. With  this representation, the eigenvalue equation
 (\ref{eq:mpsi=epsi})   becomes
\begin{eqnarray}
 &&  E \psi(x_1,\dots, x_n) =   \nonumber \\   &&
  \sum_i  p \left[ 
             \psi(x_1, \dots, x_{i-1},\ x_i-1,\ x_{i+1}, \dots, x_n) 
 - \psi(x_1,\dots, x_n) \right] + 
    \nonumber \\      
  &&  \sum_j  q \left[ \psi(x_1, \dots, x_{j-1},\ x_j+1,\ x_{j+1}, \dots, x_n)
              - \psi(x_1,\dots, x_n) \right] \, , 
\end{eqnarray}
where the sum runs over the indexes $i$ such that $ x_{i-1} < x_i-1$
 and   over the indexes $j$ such that $  x_j + 1 < x_{j+1} \,,$   {\it
i.e.}, such that the corresponding jumps  are  allowed.

\subsection{Translation invariance}
 \label{sec:Tinv}

 The translation operator $T$ shifts  simultaneously  all the particles 
  one site forward: 
\begin{equation}
   T (\tau_1, \tau_2, \ldots, \tau_L)  = 
 (\tau_L, \tau_1, \tau_2, \ldots, \tau_{L-1})    \, .
 \label{eq:defT}
\end{equation}
Because of  periodicity  we have $T^L = 1$.
  Thus,  the eigenvalues (or impulsions)  $k$ of $T$
 are simply  the $L$-th roots of unity:
 \begin{equation}
  k  =  e^{2i\pi m/L} \ \ \, \hbox{ for } m = 0, \ldots, L -1 \, .
\label{eq:eigenvalT}
\end{equation}
 We denote by  ${\bf T}(k)$ the 
 eigenspace of $T$  corresponding  to  the impulsion $k$.
 The projection  operator $\mathcal{T}_k$  over  this eigenspace is given by 
 \begin{equation}
  \mathcal{T}_k = \frac{1}{L}  \sum_{r=0}^{L-1}  k^{-r} T^r  \, .
\label{eq:projector}
\end{equation}
 The complex conjugation transforms a $T$-eigenvector  with
 eigenvalue $k$ into an eigenvector  with
 eigenvalue $k^* = k^{-1}$, {\it i.e.},  ${\bf T}(k^*) = [{\bf T}(k)]^*$.

 The ASEP on a periodic lattice is translation invariant, {\it i.e.},
 \begin{equation}
  M  T  = T M  \, .
 \label{eq:MT=TM}
\end{equation}
The  matrix $M$  and the translation operator $T$ can therefore
 be simultaneously diagonalized:  $M$  
 leaves  each  eigenspace ${\bf T}(k)$  invariant.  We denote
 by ${\bf Sp}(k)$  the 
 set of the eigenvalues of  $M$ restricted to ${\bf T}(k)$.
 Using  complex conjugation we obtain 
 the property
 \begin{equation}
       {\bf Sp}(k^*) =   \left[{\bf Sp}(k) \right]^*   \, .
\label{eq:Spconj}
\end{equation}

\subsection{Right-Left Reflection}

  The  reflection operator $R$  reverses   the right and the left 
 and  is defined by 
\begin{equation}
   R (\tau_1, \tau_2, \ldots, \tau_L) 
  =  (\tau_L, \tau_{L-1}, \ldots, \tau_2, \tau_1)           \, .
 \label{eq:defR}
\end{equation}
 We have $R^2 = 1$;  the eigenvalues of $R$
 are thus $r = \pm 1$. The reflection  $R$ reverses
   the translations,  {\it i.e.},
  \begin{equation}
 RT = T^{-1} R  \, ,
 \label{eq:RT}
\end{equation}
 and transforms a $T$-eigenvector  with
 eigenvalue $k$ into an eigenvector  with
 eigenvalue $k^* = k$.
  The operator  $R$ does not commute with the Markov matrix $M$ because 
  the  asymmetric jump  rates  $(p \neq q)$ are  not invariant 
 under  the  exchange  of  right and left.
 Writing explicitly
 the dependence of $M$ on the jump rates, we have 
\begin{equation}
  R M_{(p, q)} R^{-1} =  M_{(q, p)}          \, .
 \label{eq:RMR}
\end{equation}
 Thus,   in general the reflection is {\it not} a symmetry of the ASEP
  (it is only a symmetry of the symmetric exclusion process).

\subsection{Charge conjugation}

  The charge conjugation operator $C$ exchanges particles and holes
 in the system, {\it i.e.}, a configuration with $n$ particles
 is mapped into  a configuration with $L - n$ particles: 
\begin{equation}
   C (\tau_1, \tau_2, \ldots, \tau_L) 
  =  ( 1- \tau_1, 1 - \tau_2, \ldots, 1 - \tau_L)             \, .
 \label{eq:defC}
\end{equation}
 The operator $C$ satisfies the relations: 
\begin{equation}
 C^2 = 1 , \, \,\,\,\,  CT = TC, \,\,\,\,\, CR = RC \, .
\label{eq:C2}
\end{equation}
 By charge conjugation,  particles 
  jumping forward 
  are  mapped  into holes  jumping forward. But holes
  jumping forward are equivalent to 
   particles jumping backward.  Writing explicitly
 the dependence of $M$  on  the number of particles
 and  on the jump rates, we have 
\begin{equation}
  C M_{(n, p, q)} C^{-1} =  M_{(L-n, q, p)}          \, .
 \label{eq:CMC}
\end{equation}
 We notice  that the number of particles is conserved by $C$ only at half
 filling ($L = 2n $). 
  Thus,   the charge conjugation is {\it not} a symmetry of the ASEP
  except  for the symmetric $(p=q)$ exclusion process at half filling.

\subsection{CR symmetry}
\label{sec:CR}

  Using   equations~(\ref{eq:RMR}) and~(\ref{eq:CMC})
 to  combine  the charge
conjugation $C$ with the reflection $R$, we obtain 
\begin{equation}
  (CR) M_{(n, p, q)} (CR)^{-1} =  M_{(L-n, p, q)}         \, .
 \label{eq:CRM}
\end{equation}
 Thus,  for a given $L$, the  $CR$ operator
  maps  the ASEP with $n$ particles into the
 ASEP with  the same jumping rates  but with $L-n$ particles.  This implies 
  that the spectrum of $M$ for $n$ particles is identical with  the spectrum
for $L-n$ particles (because  $CR$ transforms eigenvectors of $M_n$ into
eigenvectors of $M_{L-n}$).

 Hereafter,  we shall always consider  the ASEP model
 at half filling, {\it i.e.}, $L
 = 2n$. In this case,  the  $CR$ operator   
 constitutes an exact symmetry because $(CR) \, M = M \, (CR)$.
 
The ASEP at half filling is therefore invariant 
under each of the two symmetries, 
translation $T$ and $CR$.  Note  that they do not commute with each
other.  Rather, we obtain from equations~(\ref{eq:RT}) and~(\ref{eq:C2})
\begin{equation}
  (CR) T = T^{-1} (CR) \, .
 \label{eq:CRT}
\end{equation}
Hence,  the $CR$   transformation  maps the subspace ${\bf T}(k)$
into ${\bf T}(k^*)$ and thus
  \begin{equation}
       {\bf Sp}(k) =   {\bf Sp}(k^*)   \, .
\label{eq:Spkk*}
\end{equation}
For $k \neq \pm 1$, the two subspaces ${\bf T}(k)$ and ${\bf T}(k^*)$ are
distinct and equation~(\ref{eq:Spkk*}) implies that the corresponding
eigenvalues of $M$ are doubly degenerate.  For $k = \pm 1$, the $CR$
transformation leaves ${\bf T}(k)$ invariant. Thus, the natural symmetries
do not predict any degeneracies in the sets ${\bf Sp}(\pm 1)$. However,
each of the subspaces ${\bf T}(\pm 1)$ is split  into two smaller
subspaces invariant under $CR$ and on which $CR = \pm 1$.

 Equations~(\ref{eq:Spconj}) and~(\ref{eq:Spkk*})  
imply that for all $k$ the set ${\bf Sp}(k)$ is 
self-conjugate,  {\it i.e.},
\begin{equation} 
 {\bf Sp}(k) = \left[  {\bf Sp}(k) \right]^*  \, .
\end{equation}
This means that ${\bf Sp}(k)$, for all $k$, is made only of 
real numbers or complex conjugate pairs.

 In the Appendix, we explain how  a  charge conjugation  operator
 $C$ can  be defined for the
 ASEP with a {\em tagged} particle  so that $CR$
 remains a symmetry.

\section{Spectral degeneracies in the TASEP}

The above discussion, based on obvious symmetries of the ASEP at half
filling with generic jump rates $p$ and $q$, suggests that the spectrum of
the Markov matrix should be composed of singlets (for impulsion $k = \pm 1$)
and doublets (for $ k \ne \pm 1$); so multiplets of higher order have no
reason to appear and therefore should not exist generically.

In this section we describe spectra for systems of size $L \le 18$ at half
filling obtained from numerical diagonalization for the TASEP, i.e. the
model where jumps are allowed  only in  one direction ($q=0$).  These spectra
contain many unexpected degeneracies of very special orders. 
 Our numerical observations lead us to conjecture  some formulae
 for general $L$. We  shall show that 
  these formulae  can be  derived,  at least 
  heuristically,  from   the Bethe Ansatz equations. 
  The existence of such degeneracies of
  higher orders strongly suggests the presence of unknown underlying
   symmetries.

\subsection{Numerical results and conjectures}

The spectral degeneracies of the Markov matrix $M$ are numerically
calculated with the following procedure.
First, the matrix $M$ is built for a given size $L$ and number of particles
$n$.  As the system is periodic, we use the translation invariance $T$
 in order to split 
the space of configurations  into $L$ subspaces ${\bf T}(k)$
with impulsion $k$ (see section  \ref{sec:Tinv}). 
 Thus  $M$ is block-diagonal with $L$ blocks. The spectrum of each block is 
 then  computed by a routine of the Lapack
library (Anderson et al., 1999) with the standard 
 numerical double precision (around 15
decimal digits).   The $L$ spectra   obtained by numerical diagonalization
are then  collected together in order to search
and count the degeneracies.

The time required to compute the complete  spectrum of a matrix of size $N$
 is  proportional to $N^3$. Since  the sizes of the blocks ${\bf T}(k)$ grow
 as  $2^L$, the computation time grows as  $8^L$. Thus  only small
systems can be studied by this method: we studied 
 systems  of size $L \le  18$. 
 
The problem is now to choose a criterion to decide when two eigenvalues, 
computed with finite precision,  are equal.  By calculating the
differences between  all the eigenvalues 
  we observe that these   differences    are  always
 either smaller than $10^{-12}$ or larger than $10^{-4}$.  Consequently
we deduce that two eigenvalues with a difference smaller than $10^{-12}$
are in fact equal and that this apparent discrepancy  is  due to 
 the finite precision of the computation.  This allows us  to enumerate the
degeneracies.
We emphasize  that  the existence of  this  gap of eight orders of magnitude 
 (between  $10^{-12}$ and $ 10^{-4}$) in 
 the data of the differences  is crucial. 
 If  the distribution of the  differences was more  continuous 
 ({\it i.e.},  if there was no obvious  gap),  the choice of a 
criterion to decide  which eigenvalues  are  identical would 
have  been much more difficult  and somehow arbitrary.

\begin{table}  \centering
  \begin{tabular}{r|r|rrrrr}
   $L$   & $\Omega$ &  $d=1$ &  $d=2$ & $d=6$ & $d=20$ & $d=70$
\\ \hline
    2    &    2     &     2 \\
    4    &    6     &     4  &     1  \\
    6    &    20    &     8  &     6  \\
    8    &    70    &    16  &    24  &    1 \\
   10    &   252    &    32  &    80  &   10 \\
   12    &   924    &    64  &   240  &   60  & 1      &     \\
   14    &  3432    &   128  &   672  &  280  & 14     &     \\
   16    & 12870    &   256  &  1792  & 1120  & 112    &  1  \\
   18    & 48620    &   512  &  4608  & 4032  & 672    & 18  \\
  \end{tabular}
  \caption{\em Spectral degeneracies in the TASEP at half filling: $L$ is
    the size of the lattice; $\Omega$ is the dimension of the configuration
    space; the other columns give $m(d)$ the number
    of multiplets  of degeneracy   $d$.
  These numbers are given by  the simple 
 combinatorial formulae~(\ref{eq:dr})
 and~(\ref{eq:mdr}).
   }
  \label{tab:num}
\end{table}

Let $m(d)$ be the number  of multiplets with degeneracy 
 of order  $d$, {\it i.e.},  the number
of sets of $d$ equal eigenvalues.  By definition, we have  $\sum_d d \ m(d) =
\Omega$, the total number of eigenvalues.  For the TASEP model with
half-filling, the numbers  $m(d)$   are given in
Table~\ref{tab:num} for sizes $L \le 18$. Although only singlets ($d=1$)
 and  doublets ($d=2$) are  predicted by  the natural
  symmetries $T$ and $CR$,  numerous degeneracies of higher order
 also   appear.  We   notice from  Table~\ref{tab:num} 
 that the observed orders of  degeneracies are only
1,2,6,20,70.  These numbers obey  the sequence
\begin{equation}
  d_r = \left(\begin{array}{c} 2r\\r \end{array} \right) 
  = \frac{ (2r)!}{r!\,  r!}
  \, ;
  \label{eq:dr}
\end{equation}
 for a system of size $2n,$  $r$ takes
 all integral values from  0  to $ n/2$.
   From  an empirical study of Table~\ref{tab:num}
 we also remark  that for $L=2n \le 18,$  
\begin{equation}
  m(d_r)  = \left(\begin{array}{c} n \\ 2r \end{array} \right) 2^{n -2r}
  \,.   
  \label{eq:mdr}
\end{equation}
  For $r = 0$ and 1, these equations provide the number  of
 singlets $ m(1) = 2^n$   and the number of doublets 
   $ m(2) = n(n-1)2^{n-3}\,. $
  We  conjecture that equations~(\ref{eq:dr})
 and~(\ref{eq:mdr}) are true for all $L$.  We have verified  that
 these  relations  satisfy  the sum rule $\sum_d d \ m(d) = \Omega$:
\begin{equation}
  \sum_{r \ge 0} d_r \ m(d_r) = 
   \sum_{0 \le r \le n/2}  \frac{  2^{n -2r} \ n!}{(n-2r)!\,  r!\,  r!}
 = \left(\begin{array}{c} 2n \\n  \end{array} \right) = \Omega \, .
  \label{eq:sr}
\end{equation}
  The last equality is obtained  by
 identifying the terms of order $x^0$ in  the identity
\begin{eqnarray}
 \left(  2 + x + \frac{1}{x}   \right)^n  = 
  \sum_{n_1 + n_2 + n_3 = n}
 \frac{ 2^{n_1} x^{ n_2 - n_3} n!   }{n_1! \,  n_2!\,  n_3!}    
  =  \frac{ (x +1)^{2n}}{x^n} =   \sum_{k}
\left(\begin{array}{c} 2n \\k  \end{array} \right)  x^{k -n}  \, .
\end{eqnarray}

  Using equations~(\ref{eq:dr})  and~(\ref{eq:mdr}), we can  analyze 
 the large $L$ (or $n$) behaviour.  For example,   we note
 that the  number of  singlets,  $2^n$, becomes  negligible 
 with respect to the total number of eigenvalues, $\Omega \sim 4^n/n^{1/2}$,
   as $L$ grows. As explained in section \ref{sec:CR}, the 
  $CR$ symmetry implies that  the singlets necessarily  belong to the 
  subspaces ${\bf T}(\pm 1)$.  However, 
  the number  of  these singlets  is much smaller 
 than the dimension of subspaces ${\bf T}(\pm 1)$ (which is approximately
 given by $\Omega / n \propto 4^n/n^{3/2}$). Therefore, multiplets
  must necessarily exist  in the subspaces ${\bf T}(\pm 1)$:
  this fact is an indication for a 
  hidden symmetry that would  commute with
  $T$ but  not commute with $CR$, for example.

 More precisely,  the number $m(d)$ of multiplets is maximal 
 for $r \sim n/6$ which corresponds to a degeneracy of order 
 $d \propto 2^{n/3}$ with  $m(d) \propto 3^n$.
 Then, we obtain that  $d \, m(d) \propto
(2^{1/3} 3)^n \ll \Omega$ because $2^{1/3} 3 \approx 3.78 < 4$.
 On the other hand,  the number $d \, m(d)$
  of eigenvalues counted with their degeneracies
  is maximal for $r \sim n/4$. In this case, we have 
  $d \propto 2^{n/2}$, $m(d)
\propto 8^{n/2}$ and $d \, m(d) \propto 4^n \propto \Omega$. Loosely speaking,
 we  deduce from this estimation that most of the  eigenvalues
 in the spectrum are degenerate with   a degeneracy of order $2^{n/2}$.


\subsection{Bethe Ansatz  analysis of the spectral degeneracies}
\label{sect:ba}

  The Markov matrix  of the exclusion process  can be diagonalized
 thanks to the {\em Bethe Ansatz}  (Dhar, 1987)  that assumes  that 
 an  eigenvector  $\psi$ of $M$  can be written as 
\begin{equation}
  \psi(x_1,\dots,x_n) = \sum_{\sigma \in \Sigma_n} A_{\sigma}  \,  
         z_{\sigma(1)}^{x_1} \,  z_{\sigma(2)}^{x_2} \dots z_{\sigma(n)}^{x_n}
  \label{eq:ba} \, , 
\end{equation}
where $\Sigma_n$ is the group of the $n!$ permutations of $n$ indexes. The
coefficients $\{A_{\sigma}\}$ and the wave-numbers $\{z_1, \dots, z_n\}$
are complex numbers to be determined.  For TASEP at half filling, it is
convenient to use the fugacity variables $Z_i = 2/z_i -1$ and to introduce
an auxiliary complex variable $Y$.  The Bethe equations can then be
reformulated as explained below.  (For more details, see Gwa and Spohn
1992; Golinelli and Mallick 2004).  In terms of the $n$-th roots of $Y$
defined as
\begin{equation}
 y_m =  Y^{1/n}  e^{(m-1)2i\pi/n} \ \ \, \hbox{ for } m = 1, \ldots, n \, , 
 \label{eq:ym} 
\end{equation}
 we  calculate    $2n$  numbers  $(Z_1, \dots, Z_{2n})$   given by 
\begin{equation}
  Z_m = (1 - y_m)^{1/2}  \,;   \ \ \ Z_{m+n} = - Z_m 
   \, .   \label{eq:zm}
\end{equation}
 In order to  select $n$ fugacities among $(Z_1, \dots, Z_{2n})$, we introduce
  a choice function $c: \{1, \dots, n\} \rightarrow
\{1, \dots, 2n\} $  that satisfies
 $  1 \le c(1) < \dots < c(n) \le 2n \, .  $ 
 The Bethe equations now   become   equivalent to the
 {\it self-consistency} equation 
\begin{eqnarray}
  A_c(Y) &=& Y  \, ,  \label{eq:ay=y} \\
  \hbox{with }  \ \ \  
   A_c(Y) &=&  -4^n \prod_{j=1}^n \frac{Z_{c(j)} - 1}{Z_{c(j)} + 1} \, . 
  \label{eq:ac}
\end{eqnarray} 
Given the choice function $c$ and a solution  $Y$ of this equation, the
$Z_{c(j)}$'s are determined from Eq.~(\ref{eq:zm}) and the corresponding
eigenvalue $E_c$ is given by  
\begin{equation}
 2E_c  =  -n + \sum_{j=1}^n Z_{c(j)} 
  \label{eq:eZ}   \, .  
\end{equation}

   We  now assume that  the 
  Bethe Ansatz is complete,  {\it i.e.},    the Bethe equations
 yield a complete basis of eigenvectors for the ASEP.  Besides,  we  
 also note  that the 
  number of different  choice functions is 
 $\Omega = (2n)! / n!^2$, $\Omega$ being precisely the 
  size of the Markov matrix.  We are thus led to  make the stronger hypothesis
  that for each choice function $c$ (among the
$\Omega$ possible choice functions), the self-consistency
equation~(\ref{eq:ay=y}) has a unique solution
  $Y$ that yields one eigenvector
$\psi_c$ and one eigenvalue $E_c$.  We further  assume that
these eigenvectors are linearly independent.

Using this one-to-one hypothesis, we will now explain the spectral
degeneracies and derive equations~({\ref{eq:dr}) and (\ref{eq:mdr}).
 We recall that  the set  $(Z_1, \dots, Z_{2n})$ is constituted of
 $n$ pairs, each pair containing  two opposite  fugacities
 $( Z_{m+n} = - Z_m )$.  A choice function $c$ selects
 $n$ numbers in the set  $(Z_1, \dots, Z_{2n})$;  amongst
 these $n$ selected fugacities, $2r$ appear in pairs
 and $n -2r$ appear as singles, with $0 \le 2r \le n$. 
 The crucial  remark is that for a given  choice function $c$
 the Bethe equations  and the  eigenvalue $E_c$
 depend only   on the $n -2r$  selected  single  fugacities. 
 This fact is clear from equations~(\ref{eq:ay=y}, \ref{eq:ac}
 and \ref{eq:eZ})
 because contributions of $Z_m$ and $Z_{m+n}$ cancel each other. 
Thus two different  choice functions $c_1$ and $c_2$ 
 that contain the same single  fugacities (but different pairs)
  lead to the same  eigenvalue but to  different eigenvectors. 
  Hence all the choice functions that contain the same 
 single  fugacities form a multiplet with the same eigenvalue.
 The number of these  choice functions is, thus, the
 order of degeneracy of the corresponding multiplet. 
  This order of degeneracy is determined solely  by the integer $r$.

  For a  given  value  of $r$ (with  $0 \le 2r \le n$),  we now calculate
  (i) the number $d_r$  of  choice functions in each multiplet,
 {\it i.e.},  its degeneracy and 
 (ii)  the  number  $m(d_r)$  of  different multiplets.

  (i)  Two  choice functions  $c$ and $c'$   are in the same multiplet
   if they 
  differ only  in the choice of the $r$  pairs. 
 The  $n -2r$  single  fugacities having been   fixed,
 we must choose $r$  pairs amongst   $2r$ available pairs. 
  The number of choice functions   in a multiplet is therefore
 given by $d_r =  \left(\begin{array}{c} 2r\\r \end{array} \right)  $
 in agreement with equation~(\ref{eq:dr}).

  (ii) A multiplet is
 characterized by  its   $s = n -2r$  single   fugacities. 
 The number of   different ways  to  select 
  these single   fugacities is given by
\begin{equation}
 m(d_r) = \frac{ 2n(2n -2)\ldots (2n - 2(s -1))}{s!} 
  = \left(\begin{array}{c} n \\ s \end{array} \right) 2^s
  = \left(\begin{array}{c} n \\ 2r \end{array} \right) 2^{n -2r}
  \,,      
\end{equation}
 in agreement with equation~(\ref{eq:mdr}).

These calculations thus explain the numerical results described in the
previous section.  However, our analysis is based on 
 the hypothesis of a one-to-one correspondence between choice functions
and linearly independent eigenvectors.
  We emphasize that this one-to-one hypothesis
  is more stringent  than the assumption of
  the completeness of the  Bethe Ansatz. 
 At  the present stage, we do not have a  proof of  this hypothesis  but 
  we have  tested  it numerically for small system sizes: it is 
   verified for $L=2, 4, 6, 8$.

\section{Conclusion and Generalizations}

 In this work, we have studied 
 the  degeneracies  in the  spectrum of the 
 Markov matrix of the totally asymmetric exclusion process
 on a periodic ring  at half filling. Numerical 
 results show the presence of  many 
 degeneracies with specific orders.  These  orders   
 and the number of multiplets of a  given  
   degeneracy can  be   calculated  from  the Bethe Ansatz
  equations  under the assumption of a one-to-one 
  correspondence   between the eigenvectors of the Markov matrix
 and some  choice functions of the fugacities that
  appear in the Bethe Ansatz. 
 Certainly, a proof (or a disproof) of this hypothesis
 would be of interest.  We believe that this  assumption must
 be true, at least in a weaker form,   because it leads to a 
  correct   enumeration of  the  spectral degeneracies  and  multiplets.
  
 We emphasize that the observed higher order degeneracies 
 are  not  predicted
 by the   natural symmetries of 
 the TASEP. Indeed, we showed that 
  invariance under  translation and  charge conjugation
 plus reflection, predicts  only  the existence of two fold degeneracies
 in the spectrum.  Therefore,  we believe that the  ASEP 
  presents some  hidden symmetries, that are responsible
  for the numerous and large  generic degeneracies in the spectrum of
  the Markov matrix. By inspecting numerically  the multiplets, we
   have observed  that each of them are composed of eigenvalues with different
  impulsion  $k$:  thus a  hidden symmetry  does not
  necessarily  commute with the  translation operator $T$.
  A  better understanding of these symmetries would be
 of  great significance.

  We have also analyzed the degeneracies in some related models 
in order to study  the influence of  asymmetry and 
 periodicity. For  the {\em partially} asymmetric exclusion process with
 non zero left and right jump rates ($p > q > 0$) on a periodic lattice and
 at half filling,   numerical computations show the existence of
  singlets and
 doublets  and the absence of
 degeneracies of higher order.  However, 
 the number of singlets is  exactly  $2^n$ ({\it i.e.}, the same as 
 for the  TASEP).  From  the  natural symmetries,  one  would 
  expect  a much larger  number of singlets
  (of the order of $4^n/n^{3/2}$). This is  again a hint for  
 some  hidden symmetries in  the partially 
  asymmetric exclusion process (Alcaraz {\it et al.} 1994). 
 The  TASEP  can be also  defined on an open lattice: it presents  
 a rich phenomenological behaviour that was  thoroughly investigated
 using  the matrix product representation (Derrida {\it et al.} 1993).
 In this system the translation symmetry $T$ is lost   and 
the number of particles is not conserved.  We have numerically
computed spectra for different  sizes $L$ and different entrance
  and exit rates:  no degeneracies 
 have been observed in any of these cases.

 Another direction of interest is to study the TASEP model for {\em
arbitrary} filling. In fact, we have numerically observed degeneracies of
high order in the spectra when the density is a simple fraction.
Degeneracies at arbitrary filling can be analyzed by Bethe Ansatz but the
calculations seem to be far more complicated.  General results will be
presented in a future work (Golinelli and Mallick 2004a).

\subsection*{Acknowledgments}

It is a pleasure to thank M.  Gaudin, V. Pasquier and S. Mallick for
inspiring discussions and remarks about the manuscript.

\appendix

\section*{Appendix : Natural symmetries of the ASEP with a tagged particle}

 To study  the motion of   particles in the  ASEP, it 
 is often interesting to tag one of them (Spohn, 1991). 
 All particles are  then  automatically labelled because overtaking is
forbidden. A configuration $\mathcal{C}$ of $n$ particles on a ring
with $L$ sites can now  be specified  by the sequence
  $(\tau_1, \tau_2, \ldots,
\tau_L; m)$ where $\tau_i = 0, 1$ according as the site $i$ is empty or
occupied and the tag is on the $m$-th occupied site 
 with $1 \le m \le n$ ($m$ is defined modulo $n$).
 The number of configurations  of the ASEP with a tagged particle 
 is   given by   $n\, \Omega = n\, L! / [ n!  (L-n)!]$.

The translation operator $T$ shifts simultaneously all the particles
one site forward, {\it i.e.}
\begin{equation}
   T(\tau_1, \tau_2, \ldots, \tau_L; m)
   = ( \tau_L, \tau_1, \ldots, \tau_{L-1}; m+\tau_L ).
   \label{eq:tm}
\end{equation}
 Indeed, the index $m$ increases  by 1 if  and only if  $T$
moves a particle from site $L$ to site 1.
   The ASEP with a tagged particle remains translation invariant:
 let $M$ be its  Markov matrix,  we have   $MT= TM$. 
  As $T^L = 1$, the eigenvalues $k$ of $T$  are   the
$L$-th roots of unity [see Eq.~(\ref{eq:eigenvalT})].

 The ASEP with   a tagged particle presents a new  symmetry $P$,
 which consists in shifting   the  tag one  particle forward 
 but  without moving the  particles, {\it i.e.} 
\begin{equation}
    P(\tau_1, \ldots, \tau_L; m)
   =  (\tau_1, \ldots, \tau_L; m+1) \, .
\end{equation}
 As $P^n=1$, the eigenvalues $\alpha$ of 
  the tag-shift operator $P$ are simply the
$n$-th roots of unity.  We have
\begin{equation}
    PM = MP ; \ \ TP = PT \ .
\end{equation}
The matrix  $M$ can  thus be simultaneously
  diagonalized with $T$ and $P$.  We call
${\bf E}(k, \alpha)$ the common eigenspace of $T$ and $P$ corresponding
 to the eigenvalues  $k$ and  $\alpha$ respectively,  where
\begin{eqnarray}
  k       &=&  e^{2i\pi j/L} \ \ \, \hbox{ for } j = 0, \ldots, L -1 \, , 
  \label{eq:km}
\\
  \alpha  &=&  e^{2i\pi a/n} \ \ \, \hbox{ for } a = 0, \ldots, n -1 \, .
  \label{eq:am}
\end{eqnarray}
and ${\bf Sp}(k, \alpha)$ the set of the eigenvalues of $M$ restricted to
${\bf E}(k, \alpha)$.

We remark  that the Markov matrix of the ASEP {\em without} a tag 
 is  the restriction of  the Markov matrix of the  ASEP with a tag
 to the subspace $\alpha = 1$. Indeed, the projection on
the subspace $\alpha = 1$ of a configuration $(\tau_1, \tau_2, \ldots,
\tau_L; m)$ is given by $ \sum_{j=1}^n (\tau_1, \tau_2, \ldots, \tau_L; j)
/ n$; this projection  renders the particles indistinguishable.

  Let us now  consider a configuration $(\tau_1, \tau_2, \ldots,
\tau_L;\ m)$, {\it i.e.},   the tagged particle 
 is the $m$-th particle from the
left.  With the left-right reflection $R$, it becomes the $m$-th one from
the {\em right}, so the $(n+1-m)$-th one from the {\em left}, {\it i.e.}, 
\begin{equation}
   R(\tau_1, \tau_2, \ldots, \tau_L;\ m)
  = (\tau_L, \ldots, \tau_2, \tau_1;\ n+1-m) \ .
\end{equation}
 We recall that $R$ is not a symmetry of the ASEP because the
 right and left jump rates, $p$ and $q$, are swapped.

 The action of the  charge conjugation operator $C$ is more subtle. 
 The operator  $C$ transforms  a system  with $n$ particles (including
 one tagged particle)
 into one with $L-n$ particles (including one tagged particle).
 The dimensions  of the corresponding configuration spaces
 will be the same  only in  the half filling case; thus 
 for the tagged ASEP, the charge conjugation  $C$
 can be  well defined only for   $L = 2n$.
  Moreover,  because  $C$ transforms  the tagged
particle into a hole,  the tag cannot remain on the same site
 and we must specify where it goes. One solution, that will  ensure
  the $CR$-invariance  of the ASEP, is to define $C$  as follows:
 \begin{equation}
   C(\tau_1, \tau_2, \ldots, \tau_L;\ m)
  = (1-\tau_1, 1-\tau_2, \ldots, 1-\tau_L;\ n+1-m) \, .
\end{equation}
 Hence, after  charge conjugation, 
 the tag  is  on  the $m$-th  particle   from the {\em right}.
Consequently, we obtain, at half filling
\begin{equation}
   CR(\tau_1, \tau_2, \ldots, \tau_L; m) 
   = (1-\tau_L, \ldots, 1-\tau_2, 1-\tau_1;m).
   \label{eq:crm}
\end{equation}
  It follows that
\begin{equation}
     (CR)^2 = 1 \ ; \ \ 
   P(CR) = (CR) P \ .
\end{equation}
It is also possible to show that $CR$ is a symmetry of the model:
\begin{equation}
   M(CR) = (CR)M \ .
\end{equation}
(a little care is needed when analyzing  jumps  between sites $L$ and 1).

Combining Eq.~(\ref{eq:tm})  with  Eq.~(\ref{eq:crm}),  
 the relation between  $CR$ and  $T$ is obtained 
\begin{eqnarray}
   CRT(\tau_1, \tau_2, \ldots, \tau_L; m)
   &=& (1-\tau_{L-1}, \ldots, 1-\tau_1, 1-\tau_L; m+\tau_L) \, ,
\\
   TCRT(\tau_1, \tau_2, \ldots, \tau_L; m)
   &=& (1-\tau_L, 1-\tau_{L-1}, \ldots, 1-\tau_1 ; m+1) \, ,
\end{eqnarray}
so $TCRT = PCR$, or equivalently
\begin{equation}
   T \, (CR) = (CR) \, P T^{-1} \, .
\end{equation}
Hence the $CR$ transformation maps the subspace ${\bf E}(k, \alpha)$ into
${\bf E}(\alpha k^*, \alpha)$ and thus 
\begin{equation}
  {\bf Sp}(k, \alpha) = {\bf Sp}(\alpha k^*, \alpha).
  \label{eq:spka}
\end{equation}
For $k^2 \ne \alpha$, these two subspaces are distinct but have  the same
spectrum: {\em their eigenvalues are doubly degenerate}.  For $k^2 =
\alpha$, the $CR$ symmetry leaves ${\bf E}(k, \alpha)$ invariant and splits
it into two smaller subspaces with $cr = \pm 1$, without degeneracy.
 With the notations of Eqs.~(\ref{eq:km}, \ref{eq:am}), the condition $k^2 =
 \alpha$ is equivalent to $j = a$ or $a+n$.  

 Complex conjugation implies that 
$ [{\bf Sp}(k, \alpha)]^* = {\bf Sp}(k^*, \alpha^*) $.  
Therefore, we obtain from 
Eq.~(\ref{eq:spka})  that ${\bf Sp}(k, \alpha)$ is self-conjugate
if  $\alpha =1 $ for any  $k$ (this case is equivalent to the ASEP without
tag) or if  $\alpha = -1$ for $k = \pm 1$. 

To summarize,  the $CR$ symmetry
can still  be defined at half filling 
 for the generic ASEP with a tagged particle. This symmetry 
 together with translation invariance  predicts
 the existence of  doublets in the spectrum [see Eq.~(\ref{eq:spka})].   

 For the totally asymmetric model (TASEP)  with a tagged particle,
numerical diagonalization  shows   numerous  degeneracies
 of order higher than 2
inside each $P$-invariant subspace.  It follows that   hidden symmetries
 should also exist in this system 
and  should commute with the tag-shift  operator $P$.

\section*{References}
\begin{itemize}

\item 
F.~C.~Alcaraz, M.~Droz, M.~Henkel, V. Rittenberg,  1994,
{\em Reaction-diffusion processes, critical dynamics,  and quantum
 chains},
 Ann. Phys.  {\bf 230}, 250.

\item 
  E. Anderson and al., 1999, 
  {\em LAPACK Users' Guide},
  (Philadelphia, SIAM)

\item 
U. Bilstein, B. Wehefritz, 1997, 
 {\em  Spectra of non-Hermitian quantum spin chains describing
 boundary induced phase transitions}, 
J. Phys. A  {\bf 30}, 4925.

\item 
 R. Bundschuh, 2002, 
{\em Asymmetric exclusion process and extremal statistics of
 random sequences},
Phys. Rev. E {\bf 65}, 031911.

\item
   D. Chowdhury, L. Santen, A. Schadschneider, 2000,
   {\em Statistical physics of vehicular traffic and some related systems},
   Phys. Rep. {\bf 329}, 199.

\item 
 B. Derrida, 1998, 
{\em An exactly soluble non-equilibrium system: the asymmetric simple
 exclusion process},  
 Phys. Rep.  {\bf 301}, 65.

\item 
 B. Derrida, M.~R.~Evans, V. Hakim, V. Pasquier, 1993,
 {\em Exact solution of a 1D  asymmetric exclusion
model  using a matrix formulation}, 
J. Phys. A: Math. Gen. {\bf 26}, 1493.

\item 
 B.~Derrida, J.~L.~Lebowitz, E.~R.~Speer, 2003,
 {\em Exact  large deviation functional  of a 
 stationary open driven diffusive system: the asymmetric exclusion process},  
 J. Stat. Phys. {\bf 110},  775.

\item 
  D. Dhar, 1987,
{\em    An exactly solved model for interfacial growth},
  Phase Transitions {\bf 9}, 51.

\item 
M. Dudzi\'nski, G.~M.~Sch\"utz, 2000,
  {\em Relaxation spectrum of the asymmetric exclusion
  process with open boundaries},
 J. Phys. A {\bf 33}, 8351.

\item 
  F.R. Gantmacher, 1964,
  {\em  Matrix theory},
  (Chelsea, New York)

\item 
O. Golinelli, K. Mallick, 2004,
 {\em Bethe Ansatz calculation of the spectral gap of the
 asymmetric  exclusion process},  
 J. Phys. A  {\bf 37}, 3321. 

\item 
 O. Golinelli, K. Mallick, 2004a,
 {\em Spectral degeneracies in the totally asymmetric exclusion process},  
 Preprint Spht T04-167.

\item 
 L.-H. Gwa, H. Spohn, 1992,
  {\em Bethe solution for the dynamical-scaling exponent of the noisy
  Burgers equation},
  Phys. Rev. A {\bf 46}, 844.

\item 
 T. Halpin-Healy, Y.-C.~Zhang, 1995, 
{\em Kinetic roughening phenomena, stochastic growth, directed polymers and
 all that},
Phys. Rep.  {\bf 254}, 215.

\item
 S.~Katz, J.~L.~Lebowitz, H.~Spohn, 1984,
  {\em  Nonequilibrium steady states of stochastic lattice gas models
 of fast ionic conductors},
 J. Stat. Phys.  {\bf 34}, 497.

\item 
  D. Kim, 1995,
 {\em Bethe Ansatz solution  for crossover scaling functions
 of the asymmetric XXZ chain  and the Kardar-Parisi-Zhang-type
 growth model},
 Phys. Rev. E {\bf 52}, 3512.

\item 
J. Krug, 1997,
 {\em Origins of scale invariance in growth processes}, 
 Adv.   Phys. {\bf 46}, 139.

\item 
 C.~T.~MacDonald, J.~H.~Gibbs, 1969,
  {\em Concerning the kinetics of poly\-peptide synthesis
 on polyribosomes}, 
Biopolymers {\bf 7}, 707.

\item
   Z. Nagy, C. Appert, L. Santen, 2002,
   {\em Relaxation times in the ASEP model using a DMRG method},
   J. Stat. Phys {\bf 109}, 623.

\item 
 P. M. Richards, 1977,
 {\em Theory of one-dimensional hopping conductivity and diffusion}, 
 Phys. Rev. B {\bf 16}, 1393.


\item
   G.M. Schutz, 2001,
   {Phase Transitions and Critical Phenomena, Vol. 19},
   (Academic, London).

\item 
H. Spohn, 1991,
{\em Large scale dynamics of interacting particles},
(Berlin, Springer).

\end{itemize}

\end{document}